\documentclass[aps,twocolumn,pra,superscriptaddress,showpacs,tightenlines]{revtex4}
\usepackage{amssymb}
\usepackage{amsmath}
\usepackage{graphicx}
\usepackage{array}
\usepackage{multirow}
\usepackage{epsfig,graphicx,times}
\usepackage{amstext}
\usepackage{latexsym}
\usepackage{bm}

\begin{document}

\title{The two-qubit controlled-phase gate based on cross-phase modulation in GaAs/AlGaAs semiconductor quantum wells}
\author{X. Q. Luo}

\affiliation{Department of Physics, Xiangtan University, Xiangtan
411105, China} \affiliation{Beijing National Laboratory for
Condensed Matter Physics, Institute of Physics, Chinese Academy of
Sciences, Beijing 100190, China}

\author{D. L. Wang}

\email{dlwang@xtu.edu.cn}

\affiliation{Department of Physics, Xiangtan University, Xiangtan
411105, China}

\author{H. Fan}
\affiliation{Beijing National Laboratory for Condensed Matter Physics, Institute of
Physics, Chinese Academy of Sciences, Beijing 100190, China}

\author{W. M. Liu}
\affiliation{Beijing National Laboratory for Condensed Matter Physics, Institute of
Physics, Chinese Academy of Sciences, Beijing 100190, China}

\date{\today }

\begin{abstract}
We present a realization of two-qubit controlled-phase
gate, based on the linear and nonlinear properties of the probe and
signal optical pulses in an asymmetric GaAs/AlGaAs double quantum
wells. It is shown that, in the presence of cross-phase modulation,
a giant cross-Kerr nonlinearity and mutually matched group
velocities of the probe and signal optical pulses can be achieved
while realizing the suppression of linear and self-Kerr optical
absorption synchronously. These characteristics serve to exhibit an
all-optical two-qubit controlled-phase gate within efficiently
controllable photon-photon entanglement by semiconductor mediation.
In addition, by using just polarizing beam splitters and half-wave
plates, we propose a practical experimental scheme to discriminate
the maximally entangled polarization state of two-qubit through
distinguishing two out of the four Bell states. This proposal
potentially enables the realization of solid states mediated
all-optical quantum computation and information processing.
\end{abstract}

\pacs{42.50.Gy, 42.65.-k, 03.67.Bg, 78.67.De}

\maketitle
%42.65.-k:nonlinear optics
%42.50.Gy:Effects of atomic coherence on propagation, absorption, and amplification of light; EIT and absorption;
%03.67.Bg:Entanglement production and manipulation
%78.67.De:Quatnum wells

\section{INTRODUCTION}

It is well known that photons are ideal carriers in all-optical quantum information processing and computation since of their potentially wide range of applications
and that they suffer little from decoherence. The lack of effective coherent photon-photon interactions and strong optical nonlinearities previously result in a serious obstacle to perform quantum computation and communication in conventional optical systems \cite{1}. Fortunately, it has already been realized that a strong enough optical nonlinearity, typically Kerr nonlinearity, which corresponds to the refractive part of third-order susceptibilities in an optical medium and
plays a pivotal role in the field of nonlinear optics, could be available to mediate a photon-photon interaction \cite{2}.

In the early days of nonlinear optics, in spite of the far-off
resonance or resonant excitation schemes, the Kerr nonlinearity is
very small or may include serious optical absorption.
Simultaneously, a large third-order susceptibilities requires the
linear susceptibility to be as small as possible for the sake of
minimizing the absorption of all fields participating in the
nonlinear process \cite{3}. However, this difficult can be solved by
introducing the electromagnetically induced transparency (EIT) in
such systems, which is capable of modifying the linear and nonlinear
optical properties of medium predominantly in the resonant atomic
systems \cite{3,4,5}. Namely, a large cross-Kerr nonlinearity with
nonabsorption have been studied theoretically and experimentally in
the atomic system through EIT or EIT related technology. For
example, Schmidt and Imamo\v{g}lu proposed a cross-phase modulation
scheme based on the EIT in a four-level atomic system \cite{6}, Kang
and Zhu experimentally observed a large Kerr nonlinearity within
vanishing linear susceptibilities in rubidium atoms \cite{7}, as
well as the possibility of highly efficient four-wave mixing and
ultra-slow optical solitons has also been explored intensively
\cite{8,9,10,11}.

Recently, it is extremely interesting to extend these quantum
coherence and interference effects to the solid state systems,
including semiconductor quantum wells (QW) and quantum dots, of
which the discrete energy levels and optical properties are
extremely analogy to atomic systems. As a consequence, there have
been numerous developments on the quantum coherence and interference
effects in semiconductor QW systems, for example, EIT and double EIT
\cite{12,13,14}, ultrafast all-optical switching \cite{15}, slow
light solitons \cite{16,17}, tunneling-induced transparency and
related phenomenon based on Fano interference
\cite{18,19,20,21,22,23,24}, etc.

As a matter of fact, it was generally recognized that such a
semiconductor QW structures also have inherent advantages such as
large electric dipole moments, high nonlinear optical coefficients,
wide adjustable parameters, and flexibility. So that it is more
helpful for realizing high-quality quantum coherence and more
promising for practical applications. In the present paper, we adopt
the GaAs/AlGaAs semiconductor QW \cite{25}, which have been realized
in the recent experiment. In our scheme, a giant cross-Kerr
nonlinearity with nearly $\pi$-conditional nonlinear phase shifts
and mutually matched group velocities (slowed) can be achieved due
to cross-phase modulation effect in the context of EIT, accompanied
by vanishing the linear and self-Kerr optical absorption. Based on
these characteristics, the two-qubit polarization quantum phase gate
can be implemented within long-time interaction and effective
maximal entanglement. Since the polarization single qubit rotation
gate can be easily realized, the universal quantum computation can
thus be achieved \cite{1}. In addition, we propose a practical
experimental scheme to identify the maximally entangled optical
polarization state of two-qubit with two out of the four Bell
states.

This paper is organized as follows. We introduce a four-level asymmetrical
coupled-double GaAs/AlGaAs semiconductor QW system with intersubband transitions in Sec. II.
The linear and nonlinear optical properties of this system are studied in Sec. III. In
Sec. IV, within the group-velocity matching of the probe and signal optical fields, we demonstrate the implementation of the
two-qubit controlled-phase gate as well as the creation of maximally entangled state
and propose a practical experiment to discriminate the maximally entangled state of the two-qubit through
discriminating two out of four Bell states. A summary of our main conclusions are given in the final section.

\section{THE ASYMMETRIC COUPLE QUANTUM WELLS}

We consider an asymmetric coupled double GaAs/AlGaAs QW composed of
four subbands (electron states) in the conduction band as shown in
Fig.1, which has been realized in the latest experiment \cite{25}.
The double QW structure consists of coupled GaAs QW of 9 and 12 nm
width, separated by 2 nm thick
$\text{Al}_{0.35}\mathrm{Ga}_{0.65}\mathrm{As}$ [see Fig. 1(a)]. For
simplicity, we consider a schematic energy diagram as Fig. 1(b)
which can also be regarded as four-level $\mathrm{N}$ configuration
system. The description of the possible transitions are dipole
allowed in such a system which interacts with two weak
linear-polarized (pulsed) probe and signal fields and a
continuous-wave (cw) pump lasers as follows. The two weak probe and
signal fields,
with half Rabi frequency ($%
\Omega_{p}=\mu _{31}E_{c}/2\hbar $) and the center frequency $\omega
_{p}$, and with half Rabi frequency $\Omega _{s}=\mu
_{42}E_{s}/2\hbar $ and the center frequency $\omega _{s}$, drive
the transition $|1\rangle \leftrightarrow |3\rangle $ and $|2\rangle
\leftrightarrow |4\rangle $, respectively. While the strong control
field with half Rabi frequency ($\Omega _{c}=\mu _{32}E_{c}/2\hbar
$) and the center frequency $\omega _{c}$ is acting on the
transition $|2\rangle \leftrightarrow |3\rangle $. Here, the dipole
moments of the transitions $|i\rangle \leftrightarrow |j\rangle $,
$\mu _{ij}(i,j=1,2,3,4)$, are the polarization unit vectors of the
laser field. The electric-field vector of
the system can be written as $E_{l}$ (l=p,c,s), likewise $\sigma ^{+}$ and $%
\sigma ^{-}$ are the unit vectors of the right-hand circularly and left-hand
circularly polarized basis, respectively.
\begin{figure}[tbp]
\includegraphics[width=3.47in]{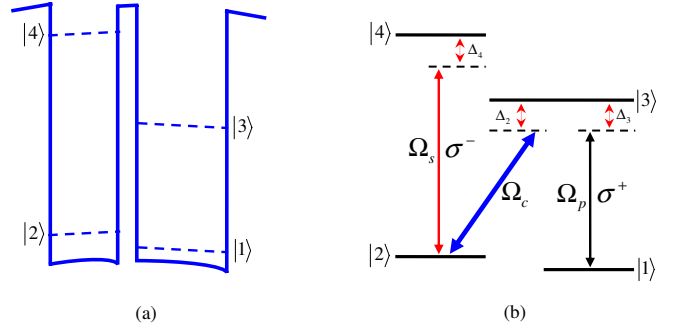}
\caption{(Color online) (a) Schematic conduction band profile for a
single period of the asymmetric double-coupled GaAs/AlGaAs quantum
wells structure consists of four discrete subband levels. (b)
Schematic energy diagram, which corresponds to (a), is consist of a
four-level quantum well system in N configuration. They are labeled
as $|1\rangle$, $|2\rangle$, $|3\rangle$ and $|4\rangle$,
respectively. $\Omega_{p}$ ($\protect\sigma^{+}$ polarization) is
the half Rabi frequency of the probe field acts on the subband transition $%
|1\rangle\leftrightarrow |3\rangle$, $\Omega_{c}$ and $\Omega_{s}$ ($\protect%
\sigma^{-}$ polarization) are the half Rabi frequency of the control and
signal fields which interact with the subband transitions $%
|2\rangle\leftrightarrow |3\rangle$ and $|2\rangle\leftrightarrow |4\rangle$
, respectively. $\Delta_{j}$ ($j$=2, 3, 4) is single-photon detunings.}
\end{figure}
\par
Since the semiconductor QW structure are low doping, the many body
effects resulting from electron-electron interactions may be
neglected in our system \cite{26}. Working in the interaction picture, by
utilizing the rotating wave approximation and electro-dipole
approximation \cite{5,9,14}, the semi-classical Hamiltonian of the system can
be written as,
\begin{equation*}
H_{int}/\hbar=-\left(
\begin{array}{cccc}
0 & 0 & \Omega _{p}^{\ast } & 0 \\
0 & \Delta _{p}-\Delta _{c} & \Omega _{c}^{\ast } & \Omega _{s}^{\ast } \\
\Omega _{p} & \Omega _{c} & \Delta _{p} & 0 \\
0 & \Omega _{s} & 0 & \Delta _{p}-\Delta _{c}+\Delta _{s}%
\end{array}
\right),
\end{equation*}
where $\Delta _{p}=\omega _{p}-\omega _{31}$, $\Delta _{c}=\omega
_{c}-\omega _{32}$ and $\Delta _{s}=\omega _{s}-\omega _{42}$ are
the one-photon detunings which denote the frequency difference
between the center and the intersubband transitions $\omega _{ij}$
$(i,j=1,2,3,4)$ of the $|i\rangle \leftrightarrow |j\rangle $. By
applying the linear Schr\"{o}dinger equation, $i\hbar\partial \Psi
/\partial t=H_{int}\Psi $, with $|\Psi \rangle $ being the
electronic energy state, the evolution equations for the subbands
probability amplitudes are
\begin{subequations}
\begin{align}
& \frac{\partial A_{1}}{\partial t}=i\Omega _{p}^{\ast }A_{3}, \\
& \frac{\partial A_{2}}{\partial t}=i(\Delta _{2}+i\gamma _{2})A_{2}+i\Omega
_{c}^{\ast }A_{3}+i\Omega _{s}^{\ast }A_{4}, \\
& \frac{\partial A_{3}}{\partial t}=i(\Delta _{3}+i\gamma _{3})A_{3}+i\Omega
_{p}A_{1}+i\Omega _{c}A_{2}, \\
& \frac{\partial A_{4}}{\partial t}=i(\Delta _{4}+i\gamma
_{4})A_{4}+i\Omega _{s}A_{2},
\end{align}
\end{subequations}
where $A_{j}$ is the probability of the subband state $|j\rangle $
(j=1-4) satisfying the conservation condition
$\sum_{l=1}^{4}|A_{l}|^{2}=1$. The detunings are defined by $\Delta
_{2}=\omega _{p}-\omega _{c}-\omega _{21}$, $\Delta _{3}=\omega
_{p}-\omega _{31}$ and $\Delta _{4}=\omega _{s}-(\omega _{p}-\omega
_{c})-\omega _{21}$ (see Fig. 1), respectively. $\gamma _{j}$
represents the decay rates of level $|j\rangle $ which results from
the the effect of lifetime broadening contribution. It is primarily
due to the longitudinal-optical phonons emission events at low
temperature, and dephasing, which is mainly owing to the
electron-electron scattering, phonons scattering processes and the
elastic interface roughness in such a QW structure.

\section{LINEAR AND NONLINEAR OPTICAL SUSCEPTIBILITIES}

To obtain the propagating properties of the probe and signal fields,
we suppose the electric field $E_{p(s)}=\varepsilon
_{p(s)}\exp[i(k_{p(s)}-\omega _{p(s)}t)]+c.c$. Under the slowly
varying amplitude approximation, we have
\begin{equation}
i\left( \frac{\partial }{\partial z}+\frac{1}{v_{g}^{j}}\frac{\partial }{%
\partial t}\right) \varepsilon _{j}+\frac{\omega _{j}}{2c}\chi
_{j}\varepsilon _{j}=0, (j=p,s)
\end{equation}
where $v_{g}^{j=p,(s)}$ is the group velocity of the probe (signal)
field,
which are defined as $v_{g}^{p(s)}=c/(1+n_{g}^{p(s)})$, with $n_{g}^{p(s)}=%
\text{Re}(\chi _{p(s)})/2+(\omega _{p(s)}/2)[\partial \text{Re}\chi
_{p(s)})/\partial \omega ]_{\omega =\omega _{p(s)}}$ being the index
of refraction and $\chi _{p(s)}$ being the susceptibilities of the
probe (signal) field.
\begin{figure}[tbp]
\includegraphics[width=3.47in,height=1.5in]{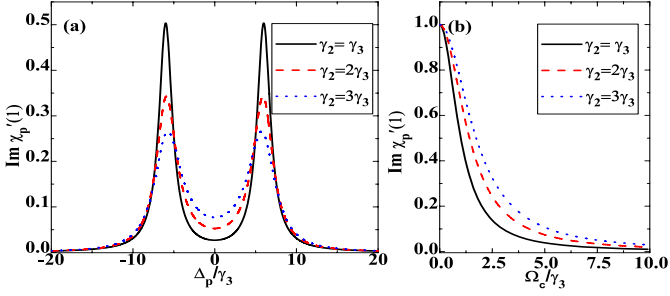}
\caption{(Color online) The imaginary part of the linear susceptibility $\protect%
\chi_{p}^{\prime(1)}$ (which represents the linear absorption
property) as a function of (a) $\Delta_{p}/\protect\gamma_{3}$ and
(b) $\Omega_{c}/\protect\gamma_{3}$ with
different decay rate $\protect\gamma_{2}$, respectively. We here make $%
\mathcal{N}\left\vert\protect\mu_{31}\right\vert^{2}/(2\protect\varepsilon%
_{0}\hbar)$ as unit in plotting. The parameters used are scaled by $\protect%
\gamma_{3}=\gamma$, and other parameters used are $\Delta_{c}=0$,
$\protect\gamma_{4}=0.5\protect\gamma$ and
$\Omega_{c}=6.0\protect\gamma$.}
\end{figure}
\par
We here suppose that the electrons are initially populated in the
subband level $\left\vert 1\right\rangle $ and the typical temporal
duration of the probe and signal fields is long enough so that the
equations can be solved adiabatically. Under these approximation, we
can obtain the steady-state solutions of Eq. (1). With the slowly
varying parts of the polarizations of the probe and signal fields
being, i.e., $P_{p}=\epsilon
_{0}\chi _{p}E_{p}=\mathcal{N}\mu _{13}A_{3}A_{1}^{\ast }$ and $%
P_{s}=\epsilon _{0}\chi _{s}E_{s}=\mathcal{N}\mu
_{24}A_{4}A_{2}^{\ast }$ (with $\epsilon _{0}$ being the
permittivity in free space and $\mathcal{N}$ being the electron
density in the conduction band of the quantum well structure ), the
expressions of the electric susceptibilities of the probe and signal
fields are
\begin{align}
\chi (\omega _{p})&=\frac{\mathcal{N}\left\vert \mu
_{13}\right\vert ^{2}}{%
2\epsilon _{0}\hbar \Omega _{p}}A_{3}A_{1}^{\ast } \nonumber\\
&\simeq \chi_{p}^{(1)}+\chi _{ps}^{(3)}\left\vert \varepsilon
_{p}\right\vert \allowbreak ^{2}+\chi _{pc}^{(3)}\left\vert
\varepsilon _{s}\right\vert \allowbreak ^{2},
\end{align}
and
\begin{equation}
\chi (\omega _{s})=\frac{\mathcal{N}\left\vert \mu _{24}\right\vert ^{2}}{%
2\hbar \epsilon _{0}\Omega _{s}}A_{4}A_{2}^{\ast }\simeq \chi
_{sc}^{(3)}\left\vert \varepsilon _{p}\right\vert \allowbreak ^{2},
\end{equation}
respectively.  Here $\chi _{p}^{(1)}$
is the linear susceptibility; $\chi _{ps}^{(3)}$ and $%
\chi _{pc}^{(3)}$ respectively depicts the third-order self-Kerr and
cross-Kerr nonlinear susceptibility of the probe field; $\chi
_{sc}^{(3)}$ denotes the cross-Kerr nonlinear susceptibility of the
signal field. The specific expressions of the linear and nonlinear
susceptibilities are shown in TABLE.I, where $d_{j}=\Delta
_{j}+i\gamma _{j}$ (j=2-4).

From TABLE.I, one finds that the probe and signal fields are
identical cross-Kerr susceptibility expressions. If the intensity of
the control field is larger than the detunings and the decay rate of
the intersubband transitions, i.e., $|\Omega_{c}|^{2} \gg
d_{2}d_{3}$, we obtain
\begin{equation}
\chi _{pc}^{(3)}=\chi _{sc}^{(3)}=-\frac{\mathcal{N}\left\vert \mu
_{13}\right\vert^{2}\left\vert \mu _{24}\right\vert^{2}}{8\varepsilon_{0}%
\hbar^{3}}\frac{1}{d_{4}\left\vert \Omega_{c}\right\vert^{2}}.
\end{equation}
\begin{table}[tbp]
\caption{ The specific expressions of linear and nonlinear
susceptibilities.} \centering
\begin{tabular}{|c|c|c|}
\hline
$\chi_{p}^{(1)}$ & $\frac{\mathcal{N}\left\vert \mu
_{13}\right\vert
                   \allowbreak ^{2}}{2\varepsilon _{0}\hbar }\chi _{p}^{\prime(1)}$
                 & $\frac{\mathcal{N}\left\vert \mu _{13}\right\vert
                  \allowbreak ^{2}}{2\varepsilon _{0}\hbar }\frac{d_{2}}{(\left\vert
                   \Omega_{c}\right\vert\allowbreak ^{2}-d_{2}d_{3})}$\\
\hline
$\chi_{ps}^{(3)}$ &$-\frac{\mathcal{N}\left\vert \mu _{13}\right\vert
                 \allowbreak ^{4}}{8\varepsilon _{0}\hbar ^{3}}\chi_{ps}^{\prime(3)}$
                &$-\frac{\mathcal{N}\left\vert \mu _{13}\right\vert \allowbreak ^{4}}{%
                  8\varepsilon _{0}\hbar ^{3}}\frac{(\left\vert \Omega _{c}\right\vert
                 \allowbreak ^{2}+\left\vert d_{2}\right\vert \allowbreak ^{2})d_{2}}{
                 (\left\vert \Omega _{c}\right\vert \allowbreak ^{2}-d_{2}d_{3})\left\vert
                 \left\vert \Omega _{c}\right\vert \allowbreak ^{2}-d_{2}d_{3}\right\vert ^{2}}$\\
\hline
$\chi_{pc}^{(3)}$ & $-\frac{\mathcal{N}\left\vert \mu _{13}\right\vert
                   \allowbreak ^{2}\left\vert \mu _{24}\right\vert \allowbreak ^{2}}{%
                    8\varepsilon _{0}\hbar ^{3}}\chi _{pc}^{\prime(3)}$
                & $-\frac{\mathcal{N}\left\vert \mu _{13}\right\vert \allowbreak
                  ^{2}\left\vert \mu _{24}\right\vert \allowbreak ^{2}}{8\varepsilon _{0}\hbar
                  ^{3}}\frac{\left\vert \Omega _{c}\right\vert \allowbreak ^{2}}{%
                  d_{4}(\left\vert \Omega _{c}\right\vert \allowbreak ^{2}-d_{2}d_{3})^{2}}$\\
\hline
$\chi_{sc}^{(3)}$ & $-\frac{\mathcal{N}\left\vert \mu _{13}\right\vert
                  \allowbreak ^{2}\left\vert \mu _{24}\right\vert \allowbreak ^{2}}{%
                  8\varepsilon _{0}\hbar ^{3}}\chi _{sc}^{\prime(3)}$
                & $-\frac{\mathcal{N}\left\vert \mu _{13}\right\vert \allowbreak
                  ^{2}\left\vert \mu _{24}\right\vert \allowbreak ^{2}}{8\varepsilon _{0}\hbar
                  ^{3}}\frac{\left\vert \Omega _{c}\right\vert \allowbreak ^{2}}{%
                 d_{4}\left\vert \left\vert \Omega _{c}\right\vert \allowbreak
                 ^{2}-d_{2}d_{3}\right\vert ^{2}}$ \\
\hline
\end{tabular}
\end{table}

In Fig. 2, we plot
the linear absorbtion property [which bases on $\chi _{p}^{\prime(1)}$] versus the various
detunings of the probe field $\Delta_{p}$ and the control field field $%
\Omega_{c}$ for different values of the decay rate $\gamma_{2}$. We
can see from Fig. 2(a) that for relatively small decay rate
$\gamma_{2}=\gamma_{3}$, a resonant probe field can propagate with
little absorption (solid curve) when the switch beam (control field $%
\Omega_{c}$) is on. Namely, in the case of
$\Delta_{p}=\Delta_{c}=0$, the probe and control fields form a
``$\Lambda$ configuration'' EIT subsystem, which induces a dark
state where the imaginary part of the linear
susceptibility $\chi_{p}^{(1)}$ vanish. When the decay rate $%
\gamma_{2}$ becomes relatively important [that is, it is larger as
shown in the dashed and dotted curves in Fig. 2(a)], it gives rise
to a resonant probe field being gradual more absorbed. This means
that, at present, the response of tunneling induced interference
between the two wells start to become less observable. Meanwhile, we
proceed to examine the linear absorption profiles corresponding to
variational Rabi frequency of control field $\Omega_{c}$ along with
the influence of the decay rate $\gamma_{2}$. We find that, in Fig.
2(b), the linear absorption coefficient
$\text{Im}\chi_{p}^{\prime(1)}$
rapidly decrease with the increase of the intensity of the control field $%
\Omega_{c}$ (see the solid curve). Simultaneously, with the slightly augment
of the decay rate $\gamma_{2}$, the decrease of the absorption coefficient $%
\text{Im}\chi_{p}^{\prime(1)}$ becomes more and more slow. Specially, under a
fixed control field $\Omega_{c}$, the absorption of the probe
correspondingly enhances with increasing the decay rate $\gamma_{2}$. It
implies that, under an appropriate condition from the contribution of the
control field $\Omega_{c}$ and the decay rate $\gamma_{2}$, the linear
absorption of the probe field can be efficiently suppressed.

In Fig. 3 we show the third-order self-Kerr nonlinear susceptibility
of the probe field [which bases on $\chi _{ps}^{\prime(3)}$] versus
the various detunings of the probe field $\Delta_{p}/\gamma_{3}$ and
the third-order cross-Kerr nonlinear susceptibilities of the probe and signal fields [which base on $\chi _{p(s)c}^{\prime(3)}$] versus the various detunings of the signal field $\Delta_{4}/\gamma_{3}$, respectively. From Fig. 3(a), one can see that, under the condition $%
\Delta_{p}=\Delta_{c}=0$, the self-Kerr nonlinear susceptibility of the
probe field can also be effectively suppressed concomitant with a
appreciably wide EIT transparent window, which is identical to the result we
have obtained in Fig. 2(a). Next, in Fig. 3(b), within the EIT transparency
window, we find that the real parts of the cross-Kerr nonlinear
susceptibilities of the probe and signal fields decay much more slowly than
the imaginary parts. Consequently, this result demonstrates that a
considerable cross phase modulation with a negligible absorption can be
created on demand by setting the signal field rather far off resonance. In
addition to this, it can also be seen that the cross-Kerr susceptibilities
of the probe and signal fields, as represented in Eq. (5), are of the same
order of the magnitudes. Therefore, these results reveal that, due to the
quantum coherence and interference between the lower subbands under the EIT
condition, not only the self-Kerr interaction can be vanished but the two
cross-Kerr susceptibilities can be enhanced predominantly.

It is clearly shown that, with suitable parameters, the corresponding
cross-Kerr susceptibilities $\chi _{pc}^{(3)}$ and $\chi _{sc}^{(3)}$ can be
predominantly enhanced in the context of EIT condition. This situation
corresponds to the case of the probe and control fields are approximate
resonance ($\Delta_{p}=\Delta_{c}=0$). Then we find, by appropriate tuning
the value of the control field $\Omega_{c}$ and the decay rate $\gamma_{4}$,
the linear absorption and self-Kerr nonlinearity susceptibilities can be
effectively suppressed, while the cross-Kerr nonlinearity susceptibility of
the probe and signal fields can be significantly enhanced.
\begin{figure}[tbp]
\includegraphics[width=3.47in]{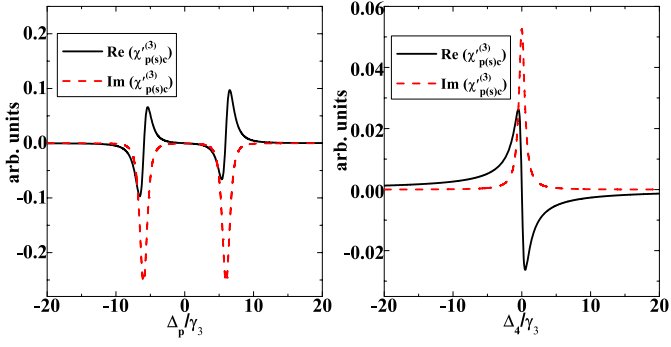}
\caption{(Color online) (a) The third-order self-Kerr susceptibility
$\protect\chi_{ps}^{^{\prime }(3)}$ versus the detunings of the
probe field $\Delta_{p}/\protect\gamma_{3}$ and (b)the third-order
cross-Kerr nonlinear susceptibilities
$\protect\chi_{p(s)c}^{^{\prime }(3)}$ of the probe (signal) field
as functions of the detunings of the signal field
$\Delta_{4}/\protect\gamma_{3}
$. We here make $\mathcal{N}\left\vert\protect\mu_{31}\right\vert^{2}/(2%
\protect\varepsilon_{0}\hbar)$ and $\mathcal{N}\left\vert\protect\mu%
_{31}\right\vert^{2}\left\vert\protect\mu_{32}\right\vert^{2}/(8\protect%
\varepsilon_{0}\hbar^{3})$ as units in plotting. The parameters used are (a) $%
\Delta_{c}=0$, $\protect\gamma_{2}=\protect\gamma_{3}=\protect\gamma$, $%
\protect\gamma_{4}=0.5\protect\gamma$, $\Delta_{4}=4.0\protect\gamma$ and $%
\Omega_{c}=6.0\protect\gamma$ and (b) $\Delta_{p}=\Delta_{c}=0$, $\protect%
\gamma_{2}=\protect\gamma_{3}=\protect\gamma$, $\protect\gamma_{4}=0.5%
\protect\gamma$, and $\Omega_{c}=6.0\protect\gamma$.}
\end{figure}

\section{GROUP-VELOCITY MATCHING AND ENTANGLEMENT OF TWO-QUBIT CONTROLLED-PHASE GATE }

It should be worth noting that, as first emphasized by Lukin and $\text{Imamo%
\v{g}lu}$ \cite{27}, the group velocity matching of the probe and signal
fields is another crucial requirement for achieving a large nonlinear mutual
phase shift. Only in this way, the two optical pulses can interact in an EIT
medium for a sufficiently long time to induce an effective cross phase
modulation.

The expressions of the group velocities for the probe and signal fields can
be given by
\begin{equation}
v_{g}^{j=p,s}\simeq\frac{4\hbar\epsilon_{0}c}{\mathcal{N}\omega_{j}}\frac{%
|\Omega_{c}|^{2}} {|\mu_{13}|^{2}(1+\beta|\Omega_{s}|^{2})\delta_{jp}+|%
\mu_{24}|^{2}\beta|\Omega_{p}|^{2}\delta_{js}},
\end{equation}
where $\beta=(\Delta_{s}^{2}-\gamma_{4}^{2})/(\Delta_{s}^{2}+%
\gamma_{4}^{2})^{2}$. And both group velocities should be small and
equal by regulating the dipole matrix elements and the probe and
signal Rabi frequencies together with the coefficient $\beta$.
Fortunately, if the EIT-resonance condition is disturbed by a small
amount, it remains under the common transparency condition and the
absorption may still be considered as ignorable.

As we know, the emergence of the cross-phase modulation is
a very essential condition for realizing the controlled quantum phase gate between two
optical qubits. In our QW scheme (as the four-state system shown in Fig. 1),
within the cross-Kerr effect, a cross-phase modulation could be implemented
whereby an optical field achieves a nonlinear phase shift relied chiefly on
the situation of another optical field to generate two-qubit quantum phase
gate. The quantum phase gate operation is defined by the input-output
relationship as $|i\rangle_{A}|j\rangle_{B}\rightarrow \exp{(i\varphi_{ij})}%
|i\rangle_{A}|j\rangle_{B}$, in which \emph{i, j}=0,1 depict the
qubit basis. In this case, a universal two-qubit gate (which enable
to entangle two initially factorized qubits) can be actualized when
the conditional phase shift $\varphi=\varphi_{00}+\varphi_{11}-\varphi_{01}-%
\varphi_{10}$ becomes different from zero \cite{1}.

Notice that only the ``right'' polarization (i.e., $\sigma^{+}$ and
$\sigma^{-}$ respectively corresponds to the probe and signal
fields) of the probe and signal fields, a controlled quantum phase
gate could be realized when a significant and nontrivial cross-phase
modulation between them arises. When the probe field has a
$\sigma^{-}$ polarization (as for a $\sigma^{-}$ polarization signal
field), namely, the ``wrong'' polarization can not couple to any
levels and thus the corresponding pulse will achieve a trivial phase
shift $ \varphi^{(0)}_{p}=k_{p}L$ (where L is the length of the
medium and $ k_{p}=\omega_{p}/\text{c}$ denotes the free space wave
vector). Under the cosideration that both the probe and the signal
fields are $\sigma^{-}$ polarization, the probe field undergoes a
self-Kerr effect and achieves a nontrivial phase shift
$\varphi_{p}^{\Lambda}=\varphi_{p}^{(0)}+
\varphi_{p}^{(1)}+\varphi_{ps}^{(3)}$ (which is subjected to the EIT
condition constituted by the $\Lambda$ configuration: $|1\rangle$,
$|2\rangle $ and $|3\rangle$ levels), with
$\varphi_{p}^{(1)}=k_{p}L(1+2\pi \chi_{p}^{(1)})$ being the linear
phase shift and $\varphi_{ps}^{(3)}$ being the nonlinear phase shift
caused by the self-Kerr nonlinearity. At the same time, the signal
field achieve the the vacuum shift $\varphi_{s}^{(0)}$. Similarly,
for the ``right'' polarization condition, the probe and signal
fields achieve the nontrivial phase shifts
$\varphi_{p}^{(T)}=\varphi_{p}^{ \Lambda}+\varphi_{pc}^{(3)}$ and
$\varphi_{s}^{(T)}=\varphi_{s}^{(0)}+ \varphi_{sc}^{(3)}$,
respectively.

The input probe and signal polarized single-photon wave packets form a
superposition of the circularly polarized states can be expressed as
\begin{equation}
|\psi\rangle_{j}=\alpha_{j}^{+}|\sigma^{+}\rangle_{j}+\alpha_{j}^{-}|%
\sigma^{-}\rangle_{j}, (j=p,s)
\end{equation}
where $|\sigma^{\pm}\rangle_{j}=\int d\omega
\xi_{j}(\omega)\alpha_{\pm}^{\dagger}(\omega)|\text{0}\rangle$, with $%
\xi_{j}(\omega)$ being a Gaussian frequency distribution of incident wave
packets, centered at frequency $\omega_{j}$. The photon field operators
experience a transformation while propagating through the QW medium of
length \emph{L}, i.e., $\alpha_{\pm}(\omega)\rightarrow\alpha_{\pm}(\omega)%
\exp\big[(i\omega/c)\int_{0}^{L}dz n_{\pm}(\omega,z)\big]$. The real part of
the refractive index $n_{\pm}(\omega,z)$ can be assumed to $%
n_{\pm}(\omega_{j},z)$, one obtains that,
\begin{equation}
|\sigma^{\pm}\rangle_{j}\longrightarrow \exp[{-i\varphi_{i}^{\pm}}]%
|\sigma^{\pm}\rangle_{j},
\end{equation}
where $\varphi_{j}^{\pm}=(\omega/c)\int_0^L\mathrm{dz}n_{\pm}(\omega_{j},z)$%
, and the cross-phase shift of the probe field is given by
\begin{equation}
\varphi_{pc}^{(3)}=k_{p}L\frac{\hbar^{2}\pi^{3/2}|\Omega_{s}|^{2}}{%
4|\mu_{24}|^{2}}\frac{\text{erf}[\zeta_{p}]}{\zeta_{p}}\text{Re}%
[\chi_{pc}^{(3)}],
\end{equation}
where $\zeta_{p}=(1-v_{g}^{p}/v{g}^{s})\sqrt{2}L/(v_{g}^{p}\tau_{s})$, with $%
\tau_{s}$ being the width of the pulse and $\text{erf}[\zeta]$
depicts the error function. The cross-phase shift of the signal
field is obtained on interchanging $\text{p}\leftrightarrow
\text{s}$ in the equation above, that is
\begin{equation}
\varphi_{sc}^{(3)}=k_{s}L\frac{\hbar^{2}\pi^{3/2}|\Omega_{p}|^{2}}{%
4|\mu_{13}|^{2}}\frac{\text{erf}[\zeta_{s}]}{\zeta_{s}}\text{Re}%
[\chi_{sc}^{(3)}],
\end{equation}
where $\zeta_{s}$ can also be obtained from $\zeta_{p}$ upon interchanging
the indices $\text{p}\leftrightarrow \text{s}$. In case of meeting with the
group velocity matching, so $\zeta_{p,s}\rightarrow 0$, i.e., the value of $%
\mathrm{erf}[\zeta_{p(s)}]/\zeta_{p(s)}$ reaches the maximum value $2/\sqrt{\pi%
}$. Thus, after encoding $|\sigma^-\rangle_i\rightarrow |0\rangle_i$ and $|\sigma^+\rangle_i\rightarrow |1\rangle_i$,
the explicit form of the polarization two-qubit controlled quantum phase gate (QPG) using the present QW structure is given by
\begin{subequations}
\begin{align}
& |0\rangle_p|0\rangle_s\rightarrow \exp\Big[-i\big(\varphi_p^{(0)}+\varphi_s^{(0)}\big)\Big]|0\rangle_p|0\rangle_s,\\
& |0\rangle_p|1\rangle_s\rightarrow \exp\Big[-i\big(\varphi_p^{(0)}+\varphi_s^{(0)}\big)\Big]|0\rangle_p|1\rangle_s,\\
& |1\rangle_p|1\rangle_s\rightarrow \exp\Big[-i\big(\varphi_p^{\Lambda}+\varphi_s^{(0)}\big)\Big]|1\rangle_p|1\rangle_s, \\
& |1\rangle_p|0\rangle_s\rightarrow\exp\Big[-i\big(\varphi_p^{(T)}+\varphi_s^{(T)}\big)\Big]|1\rangle_p|0\rangle_s,
\end{align}
\end{subequations}
with a conditional phase shift $\varphi^\text{con}=\varphi_p^{(T)}+
\varphi_s^{(T)}-\varphi_p^{\Lambda}-\varphi_s^{(0)}$ is nonzero (on the
basis of the EIT condition, the linear effect and self-phase modulation can
be suppressed, so that $\varphi_p^{\Lambda}=\varphi_p^{(0)}$, $\varphi^\mathrm{con}=\varphi_{pc}^{(3)} +\varphi_{sc}^{(3)}$),
and thus a two-qubit polarization controlled quantum phase gate can be realized in the coupled QW structure.

To show explicitly the form of the controlled phase gate, we next study the entanglement of the
two-qubit state after the processing of EIT.
For our special two-qubit system, the state takes the form $|\psi^{ps}\rangle$
 constituting by signal and probe fields. Consider the input state is a factorized (separable)
 pure state $|\psi^{ps}\rangle =|\psi \rangle _p|\psi \rangle _s$ as presented in Eq.(7),
 while the signal and probe fields $|\psi \rangle _{p(s)}$ are superposition of $|0\rangle $ and
 $|1\rangle $ with equal amplitudes $|\psi \rangle =(|0\rangle +|1\rangle )/\sqrt {2}$.
 Then by the transformation presented in Eqs.(11), the final state is still a pure state but
 with a non-trivial controlled phase so that the final state is entangled. The
 entanglement can be measured by the von Neumann entropy of the reduced density operator
 of signal field or probe field
\begin{align}
\mathrm{E}(|\psi^{ps}\rangle)&=S(p)=-\mathrm{tr}[\rho_p\log\rho_p]\nonumber\\
&=S(s)=-\mathrm{tr}[\rho_s\log\rho_s],
\end{align}
here, $\rho_p$ and $\rho_s$ are the reduced density operators of the subsystems $S^p$ and $S^s$.
Simply, if $\lambda_x$ are the eigenvalues of $\rho_{p(s)}$ [1], the entanglement of two-qubit
 expressed as von Neumann entropy takes the form,
  $\mathrm{E}(|\psi^{ps}\rangle)=-\sum_x \lambda_x\log\lambda_x$.

\begin{figure}[tbp]
\includegraphics[scale=0.8]{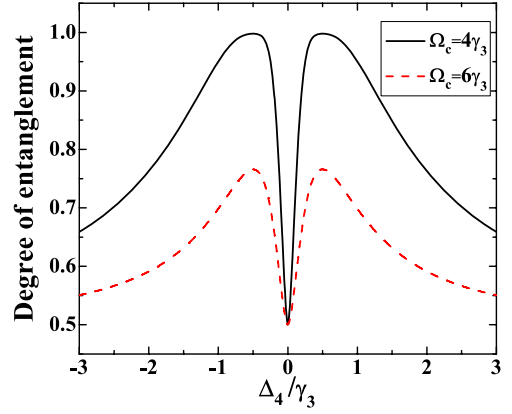}
\caption{(Color online) The degree of entanglement versus the
detunings of the signal field $\Delta_{4}/\protect\gamma_3$ with
different $\Omega_c$. With parameters $\Omega_p =\Omega_s
=0.5\protect\gamma$, $L=7 \mu m$, the others parameters used are the
same as Fig. 2}
\end{figure}

As is shown in Fig. 4, we plot the result of the degree of entanglement [von Neumann entropy $\mathrm{E}(|\psi^{ps}\rangle)$] versus the detunings of the signal field
$\Delta_{4}/\protect\gamma_3$ with different $\Omega_c$. For simplicity, we set the value of $\mathcal{N}%
\left\vert \mu _{13}\right\vert^{2}\omega_p/2\varepsilon_{0}c\hbar =\mathcal{%
N}\left\vert \mu _{24}\right\vert^{2}\omega_s/2\varepsilon_{0}c\hbar
=2.5\times10^6$ as an example. For a certain intensity of control field, i.e., $\Omega_c =4\gamma$, we can obtain a nearly 100\% degree of entanglement when $\Delta_4=\pm0.5\gamma_3$ (which corresponds to the black solid curve in
Fig. 4), which is due to the characteristic of nonabsorption in this system. In accordance with the case of the maximum entanglement, we simultaneously achieve an approximate $\pi$ radians conditional nonlinear phase shift (that is $\varphi^{con}\simeq\pi$). It is also essential to note that the degree of entanglement of our system can be affected by the fluctuations of light intensities and the detunings of the probe and signal fields in the experimental demonstration.

It is evidently shown that, a strong third-order cross-Kerr
nonlinearity is satisfied so that one can be capable of constructing
a controlled-$\pi$, a specific case of the controlled-phase gate for
the probe and signal fields in the case of the ``right''
polarization configuration. As mentioned above, it shifts the phase
of $|1\rangle_p|0\rangle_s$ by $\pi$, leaving the other three basis
states unchanged. Consequently by single qubit rotation of signal
and probe fields, the controlled-$\pi$ transformation realized by
the output modes of Eqs.(11),
 can be represented as $|0\rangle_p|0\rangle_s\rightarrow|0\rangle_p|0\rangle_s$, $|0\rangle_p|1\rangle_s\rightarrow |0\rangle_p|1\rangle_s$, $|1\rangle_p|0\rangle_s\rightarrow|1\rangle_p|0\rangle_s$ and $|1\rangle_p|1\rangle_s\rightarrow-|1\rangle_p|1\rangle_s$,
 which is the standard form in quantum computation. To our knowledge, this result indicates that,
 when reasonable regulation parameters are selected, this controlled-phase gate can
 create a maximally entangled state from a separable pure input state, (100\% entanglement as shown in Fig. 4).

Since the single qubit rotation in Eq. (7) can be realized easily,
the combination of controlled phase gate provided by EIT in the
semiconductor system and the single rotation gate constitutes a
universal set of gates for quantum computation.

\begin{figure}[tbp]
\includegraphics[scale=0.6]{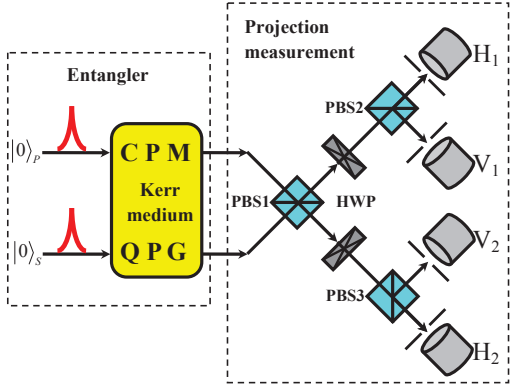}
\caption{(Color online) Scheme of the proposed experiment for a
Bell-states analyzer: CPM is the cross-phase modulation of the probe
and signal fields
in our system with conditional nonlinear phase shift $\protect\varphi^{con}=%
\protect\pi$ under the condition of EIT effect; PBS1,PBS2 and PBS3
are three polarization beam splitters; and $|H\rangle$ and
$|V\rangle$ denote the horizontally and vertically polarized
signal-photon states .}
\end{figure}
\par

The experimental scheme to realize the controlled quantum phase gate
in the QW in GaAs/AlGaAs semiconductor system is thus finished.
Next, it will be interesting to check whether this scheme can indeed
work, in particular, whether a maximally entangled state can be
created by this controlled-$\pi$ phase gate. Thus it is desirable to
have a scheme for observing the maximally entangled state among the
two-qubit polarization controlled quantum phase gate. We here
propose a practical experiment implementation for optical Bell-state
analyzer in the coincidence basis, due to which one can also observe
the ``right'' polarization and maximally entangled qubit. In order
to explain our projection, as shown in Fig. 5, we first examine that
the left part of the scheme is comprised of quantum phase gate which
has been described above. The right part of the scheme is comprised
of three polarizing beam splitters (PBS), two half-wave plate and
four detectors with single-photon sensitivity. We here encoded
$|\sigma^-\rangle_i \rightarrow |H\rangle_i$ and $
|\sigma^+\rangle_i \rightarrow |V\rangle_i$. Due to the PBS
transmits only the horizontal polarization component and reflects
the vertical component, and the input of PBS1 is the
indistinguishable identical particles, we can directly achieve the
incident state as
\begin{equation}
|\psi_{in}\rangle=a|H_1\rangle|H_2\rangle+b|H_1
\rangle|V_2\rangle+c|V_1\rangle|H_2\rangle+d|V_1\rangle|V_2\rangle,
\end{equation}
where the tensor product of the single-photon polarization basis
states, as usual $ |H\rangle_i=\Big({{1\atop 0}}\Big)_i$ and
$|V\rangle_i=\Big({{0\atop 1}} \Big)_i$. It is also well known that
the four Bell states are,
$|\phi^{\pm}\rangle=\Big(|V_1\rangle|V_2\rangle\pm|H_1\rangle|H_2
\rangle\Big)/\sqrt{2}$ and
$|\Phi^{\pm}\rangle=\Big(|V_1\rangle|H_2\rangle
\pm|H_1\rangle|V_2\rangle\Big)/\sqrt{2}$. According to the spirit of
reference \cite{29}, by using the coincidence between the four
detectors. Meanwhile, we can readily identify $|\psi_e\rangle$ in
terms of Bell states
\begin{equation}
|\psi_e\rangle=\Big[(a+d)\phi^{+}+(a-d)\Phi^{-}\Big]/\sqrt{2},
\end{equation}
where the states $\phi^+$ and $\phi^-$ are ambiguously distinguishable (as reference [30],
we can correspond $|\phi^+\rangle$ and $|\phi^-\rangle$ to $|H_1\rangle|H_2\rangle$ and $|V_1\rangle|H_2\rangle$, respectively),
thus, a half-wave plate was oriented at $\pi/8$ rotated the polarization of the horizontally
polarized single photon to $\pi/4$ polarization state just before PBS2 and PBS3, respectively.
Namely, $|H_1\rangle\rightarrow\Big(|H_i\rangle+|V_i\rangle\Big)/\sqrt{2}$ and
$|V_i\rangle\rightarrow\Big(|H_i\rangle-|V_i\rangle\Big)/\sqrt{2}$, where (i=1,2).
Finally, each of the two polarizations are split by PBS2 and PBS3 into two single photon detectors, respectively.
And thus $\phi^+$ and $\phi^-$ will be transformed into
\begin{align}
|\phi^+\rangle\rightarrow|\phi^+\rangle=\Big(|H_1\rangle|H_2\rangle+|V_1\rangle|V_2\rangle\Big)/\sqrt{2},\\
|\phi^-\rangle\rightarrow|\Phi^+\rangle=\Big(|H_1\rangle|V_2\rangle+|V_1\rangle|H_2\rangle\Big)/\sqrt{2}.
\end{align}
This shows that, we are able to identify two of the four incident Bell states using the
coincidence between the polarization state of the probe and signal fields.
Correspondingly, for the other two incident Bell states, which will result
in no coincidence between the four detectors and will be signified by the
kind of superposition of $|H_p\rangle|V_s\rangle$ and $|V_p\rangle|H_s\rangle
$. Specially, it is clear that we can also identify the maximally entangled state among the
two-qubit polarization quantum phase gate, based on the Bell state analyzer in this QW structures.

\section{DISCUSSION AND CONCLUSION}

It is important to emphasize that the measurement
of total nonlinear phase shift is crucial to the experimental
demonstration of the quantum phase gate. The fluctuations of light
intensities and frequency detunings of the probe and signal fields will
induce the errors of the nonlinear phase shift. As a consequence, there is
certainly the possibility of taking all lasers phase-locked to each other
for minimizing the effect of relative detuning fluctuations. And then the
light intensity with fluctuations of 1\% will lead to an error less than 4\%
in the phase measurement. Besides, it should also be noted that, for the
moderate density even at room temperature, the additional broadening effects
(which are induced by the carrier-carrier and carrier-photon interactions)
we have neglected are very small in comparison with the final broadening
[26, 31]. Moreover, it should be pointed out that the cross-phase modulation
is a very promising candidate for the design of deterministic optical
controlled quantum phase gates and the probe and signal fields have been treated in a
classical way. Therefore, it would be a clear indication for generating the
entanglement of macroscopic, coherent states instead of single-photon states.

In summary, we have investigated the two-qubit controlled-phase gate
on the basis of the linear and nonlinear properties of the probe and
signal pulses in an asymmetric AlGaAs/GaAs coupled-double QW
structure. In our scheme, a giant cross-Kerr nonlinearity (which
corresponds to $\pi$ radians conditional nonlinear phase shifts) and
mutually matched (which is also slow) group velocities can be
achieved within the suppression of both the linear and self-Kerr
nonlinear optical absorption susceptibilities. Such properties stems
from a constructive quantum interference effect in nonlinear
susceptibility of the probe and signal fields related to the
cross-phase modulation induced by EIT effect in our system. Due to
such novel features, we can be able to acquire two-qubit
controlled-phase gates and high entanglement between the weak probe
and signal pulses in the given system. In the mean time, it is also
likely to achieve the controlled-$\pi$ gate through the optical
realization of the circuit in quantum computation. Furthermore, by
adding just polarizing beam splitters and half-wave plates, we have
proposed a practical experimental scheme, by which it comes in handy
to discriminate the maximally entangled state of the two-qubit
including two out of four Bell states by using the coincidence
between the four detectors. And that such version of the Bell state
analyzer can possibly be extended to the three-particle or
N-particle cases. Considering that an asymmetric AlGaAs/GaAs
coupled-double QW structure on the basis of the intersubband
transitions has already be realized experimentally \cite{25}, the
results achieved in the presented work are helpful for facilitating
actual applications of all-optical quantum computing and quantum
information processing mediated by a solid-state system.

\begin{acknowledgments}
This work was supported by the NKBRSFC under grants Nos. 2011CB921502, 2012CB821305, 2009CB930701, 2010CB922904, 2011AA050526, and NSFC under grants Nos. 10934010, 11175248, 11074212, 51032002, 60978019, and NSFC-RGC under grants Nos. 11061160490 and 1386-N-HKU748/10.
\end{acknowledgments}

\end{document}